\documentclass[12pt,a4paper]{article}
\usepackage[dvips]{graphicx}
\usepackage{xspace}
\usepackage{epsfig}
\usepackage{cite}
\usepackage{axodraw}
\newcommand\smfrac[2]{{\textstyle{\frac{#1}{#2}}}}
\newcommand{\fa   }{\ensuremath{F_{\mathrm{A}}^{\gamma}}\xspace}
\newcommand{\fb   }{\ensuremath{F_{\mathrm{B}}^{\gamma}}\xspace}
\newcommand{\fl   }{\ensuremath{F_{\mathrm{L}}^{\gamma}}\xspace}
\newcommand{\fT   }{\ensuremath{F_{\mathrm{T}}^{\gamma}}\xspace}
\newcommand{\ft   }{\ensuremath{F_{\mathrm{2}}^{\gamma}}\xspace}
\newcommand{\FtA  }{\ensuremath{\widetilde{F}_{\mathrm{A}}^{\gamma}}\xspace}
\newcommand{\FtB  }{\ensuremath{\widetilde{F}_{\mathrm{B}}^{\gamma}}\xspace}
\newcommand{\FtL  }{\ensuremath{\widetilde{F}_{\mathrm{L}}^{\gamma}}\xspace}
\newcommand{\FtT  }{\ensuremath{\widetilde{F}_{\mathrm{T}}^{\gamma}}\xspace}
\newcommand{\Ftt  }{\ensuremath{\widetilde{F}_{\mathrm{2}}^{\gamma}}\xspace}
\newcommand{\der  }{\ensuremath{{\mathrm d}}\xspace}
\newcommand{\ef   }{\ensuremath{{\rm e_{\rm f}}}\xspace}
\newcommand{\efv  }{\ensuremath{\ef^4}\xspace}
\newcommand{\mf   }{\ensuremath{m_{\rm f}}\xspace}
\newcommand{\pf   }{\ensuremath{p_{\rm f}}\xspace}
\newcommand{\qsq  }{\ensuremath{Q^{2}}\xspace}
\newcommand{\psq  }{\ensuremath{P^{2}}\xspace}
\newcommand{\wsq  }{\ensuremath{W^{2}}\xspace}
\newcommand{\az   }{\ensuremath{\chi}\xspace}
\newcommand{\ts   }{\ensuremath{\theta^\star}\xspace}
\newcommand{\cts  }{\ensuremath{\cos\ts}\xspace}
\newcommand{\gevsq}{\ensuremath{{\rm GeV^2}}\xspace}
\newcommand{\gsg  }{\ensuremath{\gamma^{\star}\gamma}\xspace}
\newcommand{\ffbar}{\ensuremath{{\rm f}\bar{\rm f}}\xspace}
\newcommand{\omfw }{\ensuremath{{\mathcal{O}}\biggl(\frac{\mf}{W}\biggr)}\xspace}
\newcommand{\omfwq}{\ensuremath{{\mathcal{O}}\biggl(\frac{\mf^2}{\wsq}\biggr)}\xspace}
\begin{document}
\begin{titlepage}
\begin{flushright}
  RAL-TR-1998-079 \\
  hep-ph/9812281
\end{flushright}
\par \vspace{10mm}
\begin{center}
{\Large \bf
  QED Structure Functions of the Photon}
\end{center}
\par \vspace{2mm}
\begin{center}
{\bf Richard Nisius}\\[5mm]
{CERN, CH 1211 Gen\`eve 23, Switzerland}\\[5mm]
{and}\\[5mm]
{\bf Michael H. Seymour}\\[5mm]
{Rutherford Appleton Laboratory, Chilton,}\\
{Didcot, Oxfordshire.  OX11 0QX.  England.}
\end{center}
\par \vspace{2mm}
\begin{center} {\large \bf Abstract} \end{center}
\begin{quote}
\pretolerance 10000
  In deep inelastic electron-photon scattering in leading order QED,
\linebreak
  ${\rm e}\gamma\to{\rm e}\gsg\to{\rm e} \ffbar$, 
  there are four non-zero structure functions.  
  We calculate them for real photons retaining the full dependence on the
  fermion mass, and show numerical results of its effect.
\end{quote}
\vspace*{\fill}
\begin{flushleft}
  RAL-TR-1998-079 \\ December 1998
\end{flushleft}
\end{titlepage}
 Leading order QED structure functions have long been calculated and can
 be found in the literature, see for example Ref.~\cite{PET-8301}. 
 However, to our knowledge the full set has never been written down in one
 place retaining full dependence on the mass of the produced fermion.
 In this paper we do that, and show numerical results.
 \par
%
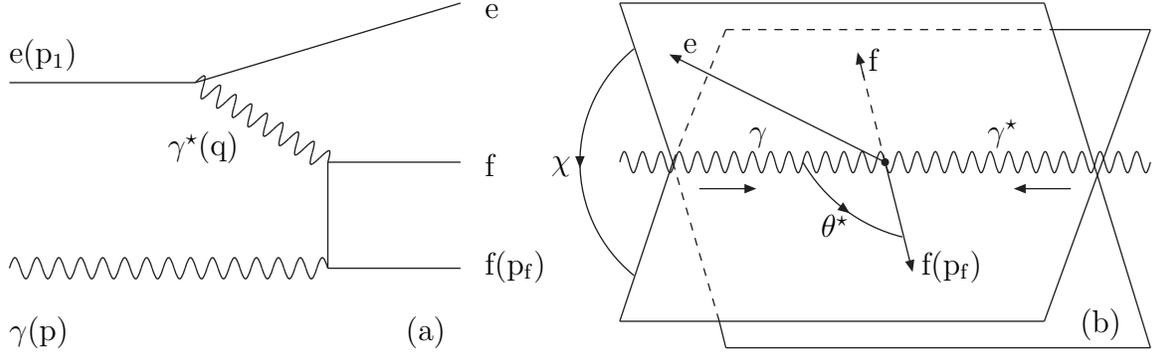
\begin{figure}[htb]\unitlength 1pt
\begin{center}
\begin{picture}(400,150)(10,50)
 \Text(150,50)[lb]{(a)}
 \Line(0,150)(70,150)
 \Line(70,150)(170,180)
 \Photon(70,150)(120,120){4}{8.5}
 \Line(120,120)(120,80)
 \Line(120,120)(170,120)
 \Text(  0,155)[lb]{$\rm e(p_1)$}
 \Text(180,175)[lb]{$\rm e$}
 \Text(180,115)[lb]{$\rm f$}
 \Text(180, 75)[lb]{$\rm f(\pf)$}
 \Text( 60,120)[lb]{$\rm \gamma^{\star}(q)$}
 \Text(  0, 50)[lb]{$\rm \gamma(p)$}
 \Photon(0,80)(120,80){4}{14.5}
 \Line(120,80)(170,80)
%
 \Text(405,53)[lb]{(b)}
 \Line(230,180)(390,180)
 \Line(395,170)(430,170)
 \DashLine(270,170)(395,170){3}
 \Line(430,50)(390,180)
 \Line(390,60)(430,170)
 \Line(230,60)(390,60)
 \Line(270,50)(430,50)
 \Line(230,180)(250,120)
 \DashLine(250,120)(267.5,60){3}
 \Line(267.5,60)(270,50)
 \DashLine(270,170)(250,120){3}
 \Line(250,120)(230,60)
 \Photon(230,120)(330,120){4}{14.5}
 \LongArrow(260,110)(280,110)
 \Vertex(330,120){1.5}
 \Photon(430,120)(330,120){4}{14.5}
 \LongArrow(400,110)(380,110)
 \LongArrow(330,120)(250,160)
 \DashLine(330,120)(320,160){3}
 \LongArrow(322,152)(320,160)
 \LongArrow(330,120)(340,80)
 \Text(370,126)[lb]{$\rm \gamma^{\star}$}
 \Text(280,126)[lb]{$\rm \gamma$}
 \Text(325,155)[lb]{${\rm f}$}
 \Text(345, 75)[lb]{${\rm f(\pf)}$}
 \Text(255,162)[lb]{${\rm e}$}
 \Text(205,115)[lb]{$\chi$}
 \ArrowArc(270,120)(55,129,231)
 \Text(307,92)[lb]{$\theta^\star$}
 \ArrowArc(351,150)(60,210,256)
\end{picture}
\caption{\label{fig:qedsf_01}
  The definition of variables used in the calculation. Shown are 
  (a) a diagram of the reaction ${\rm e}\gamma\to{\rm e}\gsg\to{\rm e}\ffbar$
  and
  (b) an illustration of the azimuthal angle \az for the reaction
  ${\rm e}\gamma\to{\rm e}\gsg\to{\rm e}\ffbar$, 
  in the \gsg centre-of-mass system.
 }
\end{center}
\end{figure}
%
 The relevant variables are $\qsq=-q^2$,
 $x=\qsq/2p\cdot q$, $y=p\cdot q / p_1\cdot p$, 
 the invariant mass squared of the fermion pair, $\wsq=(p+q)^2$ and
 $z=p\cdot \pf /p\cdot q$. 
 They are defined from the four-vectors in Figure~\ref{fig:qedsf_01}(a).
 The variable $z$ is related to the fermion scattering angle \ts in the 
 \gsg centre-of-mass frame, via $z=\smfrac12(1+\beta\cts)$, with
 $\beta=\sqrt{1-4\mf^2/W^2}$, where \mf denotes the mass of the fermion.
 The azimuthal angle \az between the electron scattering plane and the 
 \ffbar plane in the \gsg centre-of-mass frame is
 defined in Figure~\ref{fig:qedsf_01}(b).
 \par
 In this paper we only consider real target photons $\psq=0$~\gevsq.
 The full differential cross section is given by:
%
\begin{equation}
  \hspace*{-5mm}
  \frac{\der\sigma}{\der x\,\der \qsq\, \der z\, \der \az/2\pi} =
  \frac{2\pi\alpha^2}{xQ^4}
  \left[1+(1-y)^2\right]
  \nonumber\\
  \left\{
  \left( 2x\FtT + \epsilon(y)\FtL \right)
  -\rho(y)\FtA\cos\az
  +\smfrac12\epsilon(y)\FtB\cos2\az
  \right\},
\end{equation}
%
 where the functions $\epsilon(y)$ and $\rho(y)$ which can be found in
 Ref.~\cite{PET-8301}, are both $1-{\mathcal{O}}(y^2)$:
\begin{equation}
 \epsilon(y)  = \frac{2(1-y)}{1+(1-y)^2} \, ,\quad\quad\quad
 \rho(y)      = \frac{(2-y)\sqrt{1-y}}{1+(1-y)^2}
\end{equation}
 and the unintegrated structure functions \FtT, \FtL, \FtA and
 \FtB are functions only of $x$, $\beta$ and $z$ (i.e. not \az):
\begin{eqnarray}
 \hspace*{-5mm}
 \FtT(x,\beta,z) &=& 
 \frac{\efv\alpha}{2\pi}\frac{1}{2z\left(1-z\right)}
 \left\{\frac{}{}
 \left[x^2+\left(1-x\right)^2\right]\left[z^2+\left(1-z\right)^2\right]
 \phantom{\frac{\left(1-x\right)^2}{z\left(1-z\right)}}\right.
 \nonumber\\ &&
 \left.\!\!\!
 +\frac{1}{2}\left(1-\beta^2\right)
  \frac{\left(1-x\right)\left[x\left(1-2z\right)^2+2z\left(1-z\right)\right]}
  {z\left(1-z\right)}
 -\frac{1}{4}\left(1-\beta^2\right)^2
  \frac{\left(1-x\right)^2}
  {z\left(1-z\right)}
 \right\}
 \hspace*{5mm}
\end{eqnarray}
\begin{eqnarray}
 \FtL(x,\beta,z) &=& 
 \frac{4\efv\alpha}{\pi}x^2\left(1-x\right)
 \left[1-\frac{1}{4}\left(1-\beta^2\right)
 \frac{1}{z\left(1-z\right)}\right]\\
 \FtA(x,\beta,z) &=& 
 \frac{4\efv\alpha}{\pi} x\left(1-2z\right) \left(1-2x\right)
 \left[1
 -\frac{1}{2}\left(1-\beta^2\right)
  \frac{1-x}{z\left(1-z\right) \left(1-2x\right)}
 \right]
 \cdot\nonumber\\ &&
 \sqrt{\left[1-\frac{1}{4}\left(1-\beta^2\right)\frac{1}{z\left(1-z\right)}
       \right]\frac{x\left(1-x\right)}{4z\left(1-z\right)}}\\
 \FtB(x,\beta,z) &=& 
 \frac{4\efv\alpha}{\pi}x^2\left(1-x\right)
 \cdot\nonumber\\ &&
 \left\{1
 +\frac{1}{4} \left(1-\beta^2\right)  \frac{1-2x}{xz\left(1-z\right)}
 -\frac{1}{16}\left(1-\beta^2\right)^2\frac{1-x}{xz^2\left(1-z\right)^2}
 \right\}
\end{eqnarray}
 Here \ef is the charge (in units of the electron charge) of the produced
 fermion.  
 These structure functions are proportional to the
 cross sections for the target photon to interact with different
 polarisation states of the virtual photon: transverse (T), longitudinal
 (L), transverse--longitudinal interference (A) and interference between
 the two transverse polarisations (B).
 Other combinations of structure functions also appear in the literature,
 most notably~\Ftt:
%
\begin{equation}
 \Ftt \equiv 2x\FtT+\FtL.
\end{equation}
%
 In the above formulae, $z$ and \az always refer to the produced
 fermion. However, we prefer to define \az slightly differently, as
 the azimuth of whichever produced particle (fermion or antifermion) has
 the smaller $z$ value, see Figure~\ref{fig:qedsf_01}(b).
 This definition leaves all the structure functions
 unchanged except that in \FtA, the factor of $(1\!-\!2z)$ is replaced
 by $|1\!-\!2z|$.
 The more well-known integrated structure functions are given by
 integrating over the kinematically allowed range in $z$, namely
 $(1\!-\!\beta)/2$ to $(1\!+\!\beta)/2$, giving:
%
\begin{eqnarray}
 \fT(x,\beta) &=& 
 \frac{\efv\alpha}{2\pi}\left\{
 \left[x^2+\left(1-x\right)^2\right]\log\left(\frac{1+\beta}{1-\beta}\right)
 -\beta 
 +4\beta x\left(1-x\right)+
 \right.
 \nonumber\\&&
 \left.
 -\left(1-\beta^2\right)\left(1-x\right)^2
 \left(\beta-\left[1-\frac{1}{2}\left(1-\beta^2\right)\right]
 \log\left(\frac{1+\beta}{1-\beta}\right)\right)
 \right\}\\
 \fl(x,\beta) &=& 
 \frac{4\efv\alpha}{\pi}x^2\left(1-x\right)
 \left[\beta-\frac{1}{2}\left(1-\beta^2\right)
 \log\left(\frac{1+\beta}{1-\beta}\right)\right]\\
 \fa(x,\beta) &=& 
 \frac{4\efv\alpha}{\pi} x\sqrt{x\left(1-x\right)} \left(1-2x\right)
 \left\{
 \beta\left[1+\left(1-\beta^2\right)\frac{1-x}{1-2x}\right]
 \right.
 \nonumber\\&&
 \left.
 +\frac{3x-2}{1-2x}\sqrt{1-\beta^2}\arccos\left(\sqrt{1-\beta^2}\right)
 \right\},\\
 \fb(x,\beta) &=& 
 \frac{4\efv\alpha}{\pi}x^2\left(1-x\right)
 \left\{\beta\left[1-\left(1-\beta^2\right)\frac{1-x}{2x}\right] 
 \phantom{\left(\frac{1+\beta}{1-\beta}\right)}
 \right.\nonumber\\
                 & &
 \left.
 +\frac{1}{2}\left(1-\beta^2\right)
 \left[\frac{1-2x}{x}-\frac{1-x}{2x}\left(1-\beta^2\right)\right]
 \log\left(\frac{1+\beta}{1-\beta}\right)\right\}
\end{eqnarray}
\pagebreak[3]
\begin{eqnarray}
 \ft(x,\beta) &=& 
 \frac{\efv\alpha}{\pi} x\left\{
 \left[x^2+\left(1-x\right)^2\right]\log\left(\frac{1+\beta}{1-\beta}\right)
 -\beta 
 +8\beta x\left(1-x\right)
 \right.
 \nonumber\\&&
 \left.
 -\beta\left(1-\beta^2\right)\left(1-x\right)^2
 \right.
 \nonumber\\&&
 \left.
 +\left(1-\beta^2\right)\left(1-x\right)
 \left[\frac{1}{2}\left(1-x\right)\left(1+\beta^2\right)-2x\right]
 \log\left(\frac{1+\beta}{1-\beta}\right)
 \right\}
\end{eqnarray}
%
 Note that if we had not redefined \az, $F_A$ would have been zero.
 In most previous analyses, for example Ref.~\cite{AUR-9601}, the small mass
 limit has been taken, neglecting terms of order $\mf/W$ and higher leading 
 to:  
%
\begin{eqnarray}
 \fT\biggl(x,\frac{\mf^2}{\wsq}\biggr) &=& 
 \frac{\efv\alpha}{2\pi}\biggl\{
 \left[x^2+\left(1-x\right)^2\right]\log\frac{\wsq}{\mf^2}
 -1 
 +4x\left(1-x\right)
 \biggr\}+\omfwq \\
 \fl\biggl(x,\frac{\mf^2}{\wsq}\biggr) &=& 
 \frac{4\efv\alpha}{\pi}\biggl\{x^2\left(1-x\right)\biggr\}+\omfwq \\
 \fa\biggl(x,\frac{\mf^2}{\wsq}\biggr) &=& 
 \frac{4\efv\alpha}{\pi}
 \biggl\{x\left(1-2x\right)\sqrt{x\left(1-x\right)}\biggr\}+\omfw \\
 \fb\biggl(x,\frac{\mf^2}{\wsq}\biggr) &=&
 \frac{4\efv\alpha}{\pi}\biggl\{x^2\left(1-x\right)\biggr\}+\omfwq
 \\
 \label{F2}
 \ft\biggl(x,\frac{\mf^2}{\wsq}\biggr) &=& 
 \frac{\efv\alpha}{\pi}x
 \biggl\{\left[x^2+\left(1-x\right)^2\right]\log\frac{\wsq}{\mf^2}-1+8x(1-x)
 \biggr\}+\omfwq
\end{eqnarray}
%
 Note however that $F_A$ has parametrically larger mass
 corrections than the other structure functions%
%
\footnote{Note also a 
  common source of confusion: if one cuts off the $z$ integration by
  introducing a cutoff in transverse momentum $p_{t,\mathrm{min}}$ one
  obtains instead of (\ref{F2}),
   \[
     F_2 = \frac{\efv\alpha}{\pi}x\left\{
           \left[x^2+\left(1-x\right)^2\right]
           \log\frac{W^2}{\mf^2+p_{t,\mathrm{min}}^2}
           - 1 + \left(6+2\frac{\mf^2}{\mf^2+p_{t,\mathrm{min}}^2}\right)x(1-x)
     \right\}
   \]
  and hence in the massless limit an identical form to (\ref{F2}) but
  with $8x(1\!-\!x)$ replaced by $6x(1\!-\!x)$,
   \[
     F_2 = \frac{\efv\alpha}{\pi}x\left\{
           \left[x^2+\left(1-x\right)^2\right]
           \log\frac{W^2}{p_{t,\mathrm{min}}^2}
           - 1 + 6x(1-x)
           \right\}
   \]
   }.
    \par
\begin{figure}
\begin{center}
{\includegraphics[width=0.90\linewidth]{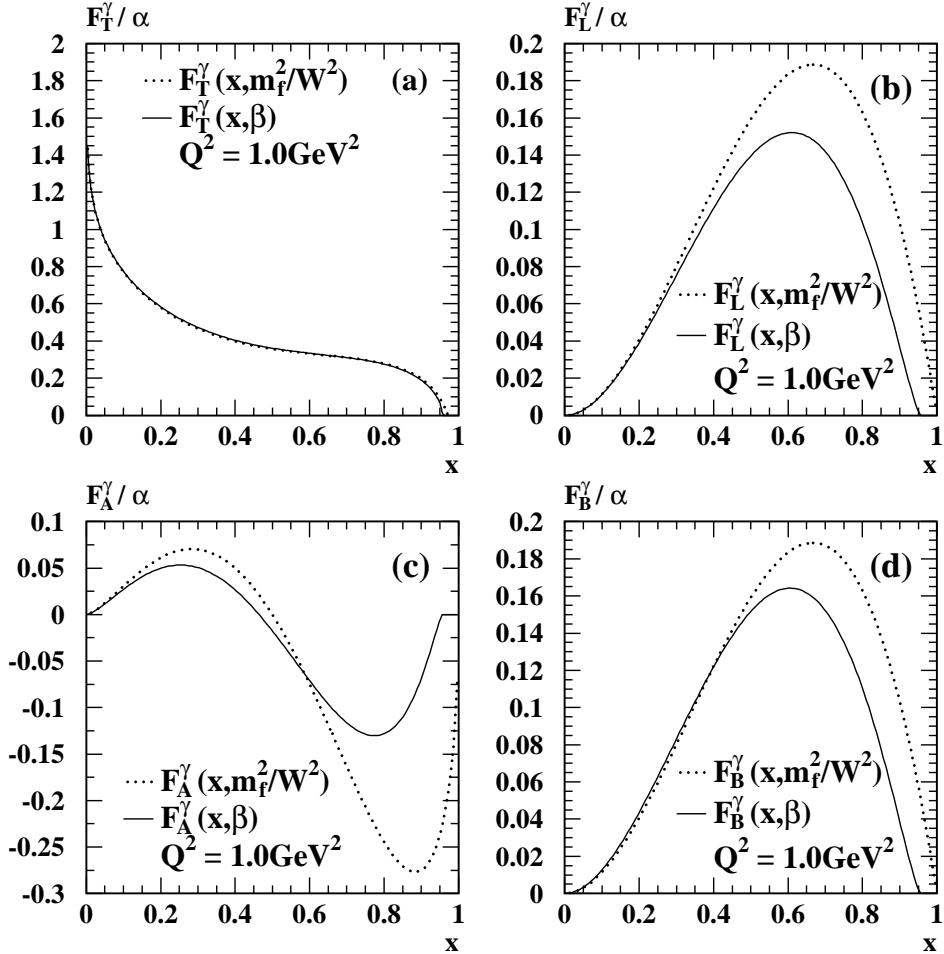}}
\caption{\label{fig:qedsf_02}
         The structure functions for $\qsq=1$~\gevsq 
         with the full mass
         dependence, $F_i^\gamma(x,\beta)$, and in the small-mass
         limit, $F_i^\gamma(x,\mf^2/W^2)$.
        }
\end{center}
\end{figure}
\begin{figure}
\begin{center}
{\includegraphics[width=0.90\linewidth]{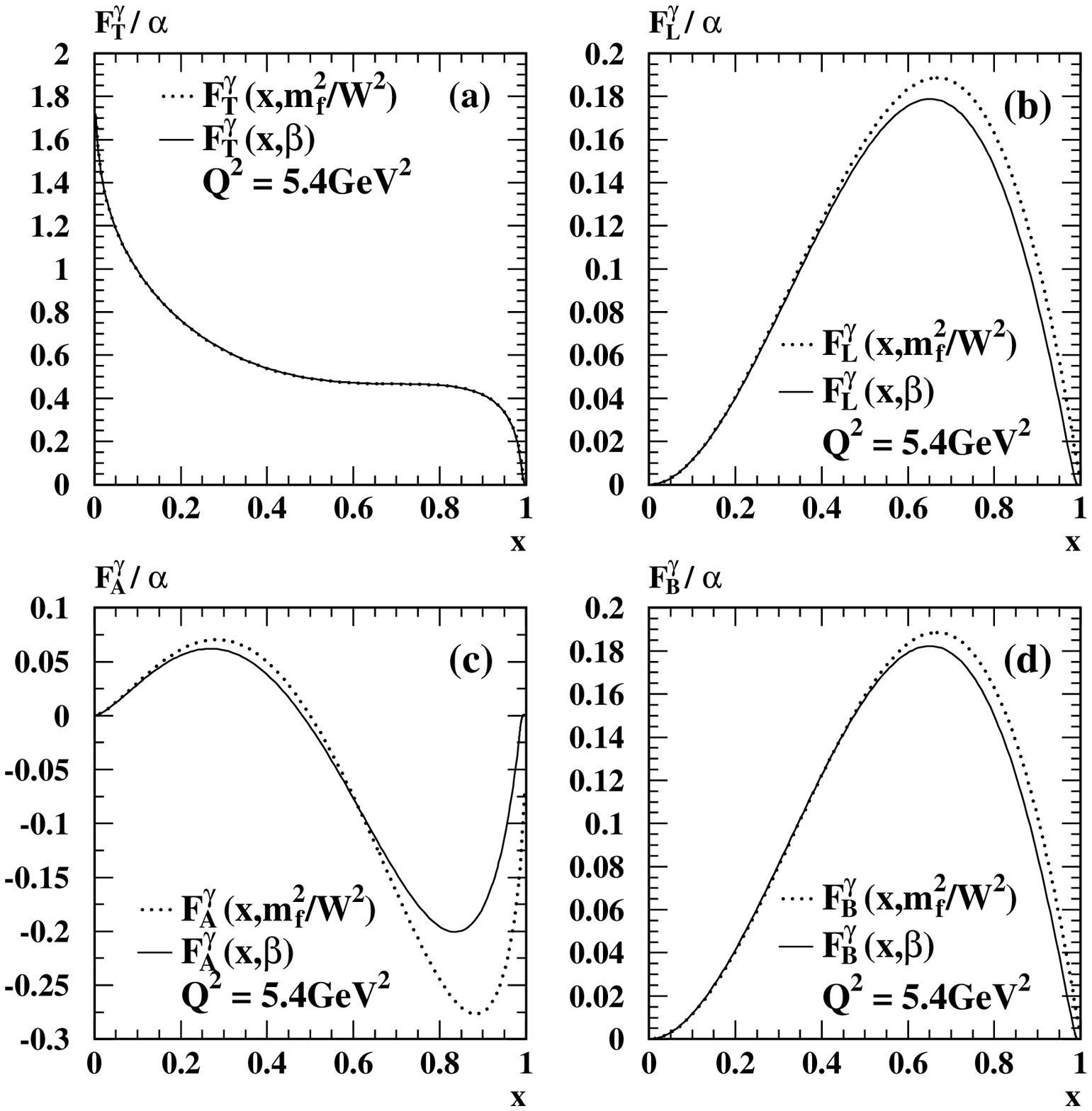}}
\caption{\label{fig:qedsf_03}
         The structure functions for $\qsq=5.4$~\gevsq
         with the full mass
         dependence, $F_i^\gamma(x,\beta)$, and in the small-mass
         limit, $F_i^\gamma(x,\mf^2/W^2)$.
        }
\end{center}
\end{figure}
\begin{figure}
\begin{center}
{\includegraphics[width=0.90\linewidth]{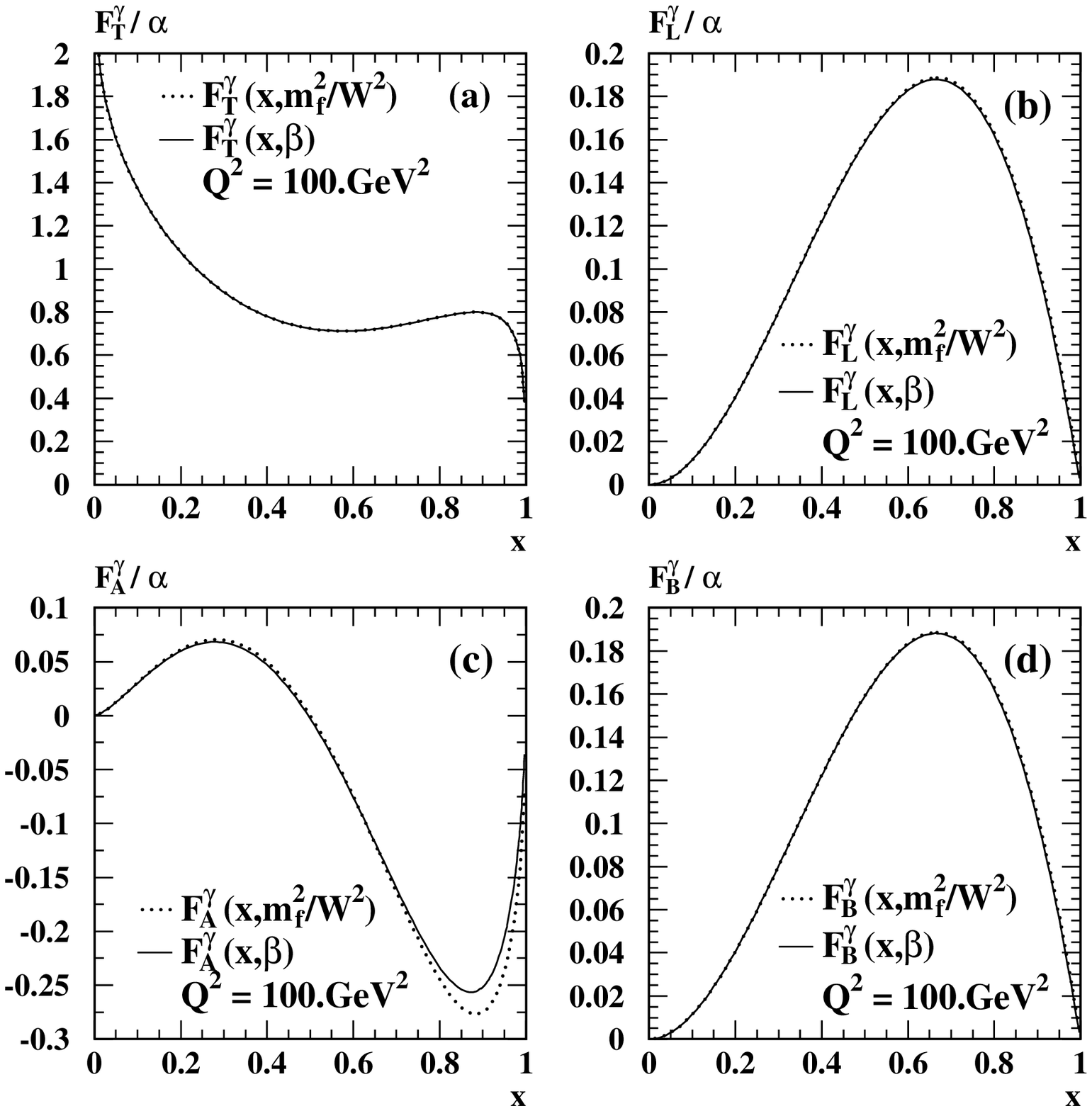}}
\caption{\label{fig:qedsf_04}
         The structure functions for $\qsq=100$~\gevsq
         with the full mass
         dependence, $F_i^\gamma(x,\beta)$, and in the small-mass
         limit, $F_i^\gamma(x,\mf^2/W^2)$.
        }
\end{center}
\end{figure}

%
 The recent experimental analysis of \fa and \fb from 
 Refs.~\cite{OPALPR182,L3C-9801} are based mainly on data in the 
 approximate range in \qsq from 1.5~\gevsq to 10~\gevsq.
 In Figure~\ref{fig:qedsf_02} we show results for the four 
 structure functions for muon final states at $Q^2=1\;\mathrm{GeV}^2$, where 
 the mass corrections are extremely important.
 Even at $Q^2=5.4\;\mathrm{GeV}^2$, Figure~\ref{fig:qedsf_03}, the mass 
 corrections for $F_A$ are sizeable but at
 $Q^2=100\;\mathrm{GeV}^2$ they are small, Figure~\ref{fig:qedsf_04}.
\clearpage
%
%

%
%
%
\end{document}